\documentclass[trackchanges,twocolumn]{aastex701}

\hypersetup{linkcolor=blue,citecolor=blue,filecolor=cyan,urlcolor=magenta}

\usepackage[utf8]{inputenc}
\DeclareUnicodeCharacter{0301}{\'{}} 

\newcommand{\h}[1]{\textcolor{black}{\textbf{#1}}}
\shorttitle{Spatially resolved spectral properties of M87* on event horizon scales}
\shortauthors{Zhao et al.}

\begin{document}

\title{Spatially resolved spectral properties of M87* on event horizon scales}

\author[0000-0002-9774-3606]{Shan-shan Zhao}
\affiliation{Shanghai Astronomical Observatory, Chinese Academy of Sciences, Shanghai 200030, P. R. China}
\email{zhaoss@shao.ac.cn}

\author[0000-0002-7692-7967]{Ru-Sen Lu*}
\affiliation{Shanghai Astronomical Observatory, Chinese Academy of Sciences, Shanghai 200030,  P. R. China}
\affiliation{Key Laboratory of Radio Astronomy and Technology, Chinese Academy of Sciences, A20 Datun Road, Chaoyang District, Beijing, 100101, P. R. China}
\affiliation{Max-Planck-Institut für Radioastronomie, Auf dem Hügel 69, D-53121 Bonn, Germany}
\email[show]{rslu@shao.ac.cn}

\author[0000-0001-7361-2460]{Rocco Lico}
\affiliation{INAF-Istituto di Radioastronomia, Via P. Gobetti 101, I-40129 Bologna, Italy}
\affiliation{Instituto de Astrof\'isica de Andaluc\'ia-CSIC, Glorieta de la Astronom\'ia s/n, E-18008 Granada, Spain}
\email{rocco.lico@inaf.it}

\author[0000-0003-0213-7628]{Yuh Tsunetoe}
\affiliation{Shanghai Astronomical Observatory, Chinese Academy of Sciences, Shanghai 200030, P. R. China}
\affiliation{Center for Computational Sciences, University of Tsukuba, \ 1-1-1 Tennodai, Tsukuba, \ Ibaraki 305-8577, Japan}

\email{ytsunetoe@shao.ac.cn}

\author[0000-0001-8492-892X]{Sijia Peng}
\affiliation{Shanghai Astronomical Observatory, Chinese Academy of Sciences, Shanghai 200030, P. R. China}
\email{sjpeng@shao.ac.cn}

\correspondingauthor{Ru-Sen Lu}

\begin{abstract}
The supermassive black hole at the center of the nearby radio galaxy M87 (M87*) is a prime target for studying black hole physics. Spatially resolved spectral measurements on event-horizon scales can reveal the origin of the emission and probe the plasma and gravitational environment in the immediate vicinity of the black hole. Here, we present an analysis of spectral properties based on nearly simultaneous high-resolution images at 3.5\,mm (86\,GHz) and 1.3\,mm (230\,GHz), obtained in 2018 with the Global Millimeter VLBI Array (GMVA) including ALMA and the Greenland Telescope, and the Event Horizon Telescope (EHT). We obtain the first spatially resolved spectral-index  map ($S_\nu \propto \nu^\alpha$) within the compact region ($\leq 100\,\mu$as). We further detect a robust radial gradient with a modest rise in the inner $\lesssim 20~\mu$as (slightly inside the 1.3\,mm ring), followed by a systematic decline at larger radii. The spectral index transitions from positive to negative values near $\sim 30~\mu$as, close to the 3.5\,mm ring radius, consistent with frequency-dependent synchrotron opacity in the innermost accretion flow. These results provide new observational constraints that can help discriminate between models of the horizon-scale emission and the launching of relativistic jets in M87*.
\end{abstract}


\section{Introduction}

The supermassive black hole (SMBH) at the center of the nearby giant elliptical galaxy M87 (M87*) provides an exceptional laboratory for probing both fundamental physics and black hole astrophysics in the immediate vicinity of an event horizon. As the first black hole imaged by the Event Horizon Telescope (EHT), M87* has become a cornerstone target for testing general relativity, studying black hole shadows, and investigating black hole spin, accretion dynamics, and relativistic jet formation~\citep{2019ApJ...875L...1E,2019ARA&A..57..467B}. Owing to its large mass and proximity \citep[\(M \sim 6.5 \times 10^9\,M_\odot\) and \(D \sim 16.8\) Mpc,][]{2011ApJ...729..119G,2019ApJ...875L...6E}, M87* subtends a sufficiently large angular scale on the sky---1\,mas corresponding to approximately 131 Schwarzschild radii (\(R_S\))---to permit horizon-scale imaging, enabling detailed tests of theory with unprecedented precision.

While previous observations have yielded important insights into the morphology and polarization structure of M87* on event-horizon scales at 1.3\,mm and 3.5\,mm \citep[e.g.,][]{2019ApJ...875L...1E,2021ApJ...910L..12E,2023ApJ...957L..20E,2023Natur.616..686L,2024A&A...681A..79E,2025arXiv250924593T}, the spatially resolved spectral properties of the emission near the event horizon remain largely unconstrained~\citep{2023A&A...673A.159R}. Such information offers essential diagnostics of the physical conditions and plasma dynamics in the immediate black hole environment, including the magnetic field, optical depth, and electron energy distribution~\citep[e.g.,][]{2023MNRAS.519.4203R}. Furthermore, characterizing the spectral dependence of the horizon-scale emission is critical for isolating the frequency-independent ``photon ring'' component \citep[e.g.,][]{Johnson2020SciA6.1310}, thereby enabling more stringent tests of general relativity.

In this work, we utilize previously published high-resolution VLBI images of M87* from observations with the Global Millimeter VLBI Array~\citep[GMVA,][]{2024evn..conf..159R} at 3.5\,mm and from the EHT observations~\citep{2019ApJ...875L...2E} at 1.3\,mm to derive spatially resolved spectral properties on event-horizon scales. Throughout the paper, the spectral index of the emitting plasma is defined as $\alpha$ in $S_\nu \propto \nu^\alpha$.

\section{Observations and Image Analysis}
The images used in this work are based on observations of M87* obtained with the GMVA in conjunction with the phased Atacama Large Millimeter/submillimeter Array (ALMA) and the Greenland Telescope (GLT; \citealt{2014RaSc...49..564I}) at a wavelength of 3.5\,mm (86\,GHz; \citealt{2023Natur.616..686L}), as well as observations from the EHT at 1.3\,mm (230\,GHz; \citealt{2024A&A...681A..79E}). The GMVA+ALMA+GLT observations were conducted on April 14–15, 2018, with data recorded at a total bandwidth of 256\,MHz per polarization. The EHT observations of M87* were carried out over four days in April 2018 (April 21, 22, 25, and 28), with data recorded in four frequency bands centered at 213.1\,GHz (band 1), 215.1\,GHz (band 2), 227.1\,GHz (band 3), and 229.1\,GHz (band 4). All stations participated in all four bands, except for the GLT, which contributed only to bands 3 and 4. Each band provided 2048\,MHz of bandwidth per polarization at each station, except for ALMA, which contributed an effective bandwidth of 1875\,MHz. Favorable weather conditions and minimal technical issues made April 21 the highest-quality observing day for M87* during the 2018 EHT campaign~\citep{2024A&A...681A..79E}.
Furthermore, since the GLT only observed in bands 3 and 4, which improved the imaging quality, we focus our analysis on the image reconstructed from the band 3 data\footnote{For simplicity, we refer to the band 3 image as the 230\,GHz image throughout this paper.}.

To obtain reliable spatially resolved spectral index measurements, we analyzed the fiducial 3.5\,mm image reconstructed with the SMILI algorithm in  \citet{2023Natur.616..686L} and the representative 1.3\,mm image published in \citet{2024A&A...681A..79E}.

\begin{enumerate}
    \item \textit{Alignment}: The 86\,GHz and 230\,GHz images were registered at the black hole center, which was determined by identifying the most probable ring center within a $25\,\mu\mathrm{as} \times 25\,\mu\mathrm{as}$ region in each image. The registration was further validated using a phase cross-correlation technique implemented in the \textit{scikit-image} package \footnote{\url{https://scikit-image.org/}} \citep{scikit-image}. 
    In this procedure, we neglected the core-shift effect associated with frequency-dependent jet emission \citep[e.g.,][]{2011Natur.477..185H}, as the observed emission at both frequencies is expected to be dominated by the accretion flow \citep{2023Natur.616..686L,2024A&A...681A..79E}. Potential systematic uncertainties associated with the image alignment are discussed in Appendix~\ref{sec:A_alignment}.
    
\item \textit{Resampling}: Both images were resampled onto a common angular pixel scale of 0.2 $\mu$as to enable direct, pixel-by-pixel comparison of the flux densities. 


\item \textit{Blurring}:
Both images were convolved with a common circular beam size to avoid resolution-induced spectral artifacts. 
We tested beam sizes ranging from $20\,\mu\mathrm{as}$, close to the achievable resolution at 230\,GHz, to $40\,\mu\mathrm{as}$, comparable to the best resolution achievable at 86\,GHz. Based on these tests, we adopted a beam size of $30\,\mu\mathrm{as}$ for the final analysis; detailed comparisons are presented in Appendix~\ref{sec:A_beam_size}.
\end{enumerate}

After applying the above processing steps, the spectral index was calculated on a pixel-by-pixel basis according to:

\begin{equation}\label{eq:alpha}
\alpha = \frac{\log(S_{\nu_2}/S_{\nu_1})}{\log(\nu_2/\nu_1)},
\end{equation}
where $\nu_1$ and $\nu_2$ are the observing frequencies, and $S_{\nu_1}$ and $S_{\nu_2}$ are the corresponding pixel-based flux densities.

While the spectral-index map provides the full two-dimensional distribution of the spectral properties, pixel-level spectral-index measurements can be influenced by azimuthal asymmetries in the 3.5\,mm image, potentially arising from non-uniform $(u,v)$ coverage \citep{2023Natur.616..686L}, as well as by day-scale structural variability between the non-simultaneous 3.5 and 1.3\,mm observations~\citep{2019ApJ...875L...4E}. To characterize the dominant radial spectral behavior while reducing sensitivity to local variations, we derive a radial spectral-index distribution from annular flux densities. Specifically, at each radius, the 86 and 230 GHz flux densities are independently integrated within a concentric annulus of width 2 $\mu$as centered on the black hole. The spectral index is then calculated from the pair of annular flux densities measured at the two frequencies. Because the flux densities are integrated over each annulus prior to calculating the spectral index, this procedure provides a flux-weighted characterization of the radial spectral properties.

In addition to the annular analysis, we derive spectral-index distributions along two orthogonal directions: one aligned with the jet axis (PA=$288^\circ$) and one perpendicular to it (PA=$18^\circ$). For each direction, flux densities at 86 and 230 GHz are integrated within consecutive 2$\mu$as$\times$2$\mu$as boxes along a narrow strip centered on the black hole. The spectral index at each position is then calculated from the integrated flux densities of the corresponding boxes. These directional distributions provide a complementary view of the spectral structure and enable a comparison between the jet and transverse directions.

The spectral index uncertainty was estimated following previous studies~\citep[e.g.,][]{2023A&A...673A.159R,2012A&A...545A.117L}:

\begin{equation}\label{eq:sigma_alpha}
\sigma_{\alpha, ij}=\frac{1}{\log(\nu_{2}/\nu_{1})}\times\sqrt   { \left(\frac {\sigma_{I_{\nu_{1}},ij}} {I_{\nu_{1},ij}}\right)^{2} +     \left(\frac{\sigma_{I_{\nu_{2}},ij}} {I_{\nu_{2},ij}}\right)^{2} },
\end{equation}
where
\begin{equation}\label{eq:sigma_I}
\sigma_{I_{\nu,ij}}=\delta_\nu I_{\nu,ij} +\sigma_{\rm{rms}\nu}.
\end{equation} 
$\delta_\nu I_{\nu,ij}$ is the systematic uncertainty, with $\delta_\nu\sim 10\%$ \citep{2023A&A...673A.159R}. $\sigma_{\mathrm{rms}\nu}$ denotes the thermal random noise, estimated from the images to be 2 mJy/beam at 86\,GHz and 1 mJy/beam at 230\,GHz.  Using these values, we find that the typical spectral index uncertainty is $\sigma_\alpha \sim 0.2$. 

For the radial and directional spectral distribution, error bars are estimated using Eqs.~\ref{eq:sigma_alpha} and ~\ref{eq:sigma_I}, where $I_{\nu,ij}$ and $\sigma_{\mathrm{rms},\nu}$ correspond to the integrated values within each annulus or box segment, respectively.
Residual systematic uncertainties associated with the amplitude calibration of the VLBI arrays and the determination of the compact flux density of the horizon-scale emission are discussed in Appendix~\ref{sec:A_flux_density}.

\section{results}

\begin{figure*}[ht!]
    \centering
   \includegraphics[width=0.9\linewidth]{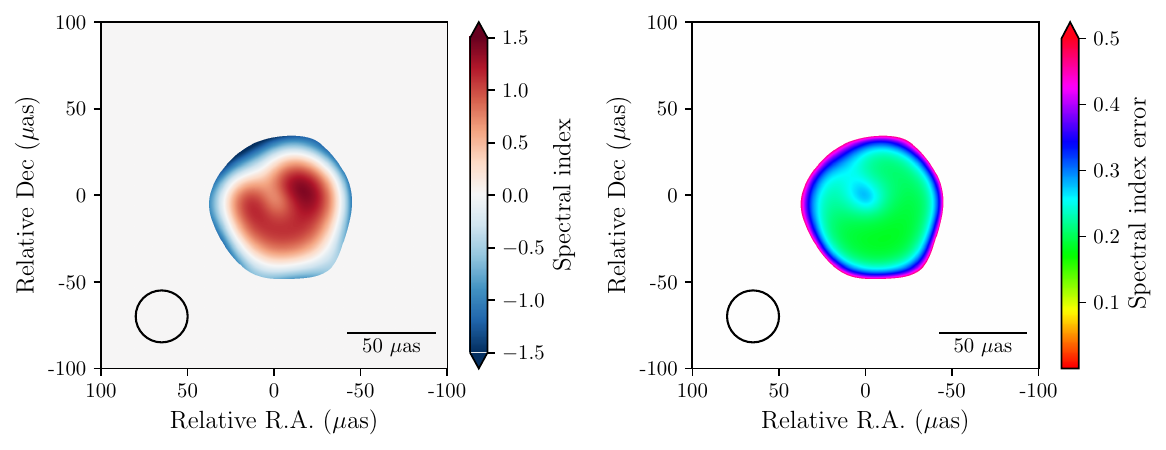}
    \caption{\textbf{Left panel}: Spatially resolved spectral index map within a $200\,\mu$as field of view, derived from horizon-scale images observed by the GMVA in 2018 \citep{2023Natur.616..686L} and the EHT in 2018 \citep{2024A&A...681A..79E}. Both images have been convolved with a $30\,\mu$as beam, indicated by the circle in the lower-left corner.
    \textbf{Right panel}: Corresponding spatial distribution of the spectral index uncertainty. In both panels, pixels are masked to include only regions where the intensity in both bands exceeds $3\,\sigma_{\rm rms,\nu}$.
    }
    \label{fig:spectral_index_map}
\end{figure*}

\begin{figure*}
    \centering
    \includegraphics[width=0.8\linewidth]{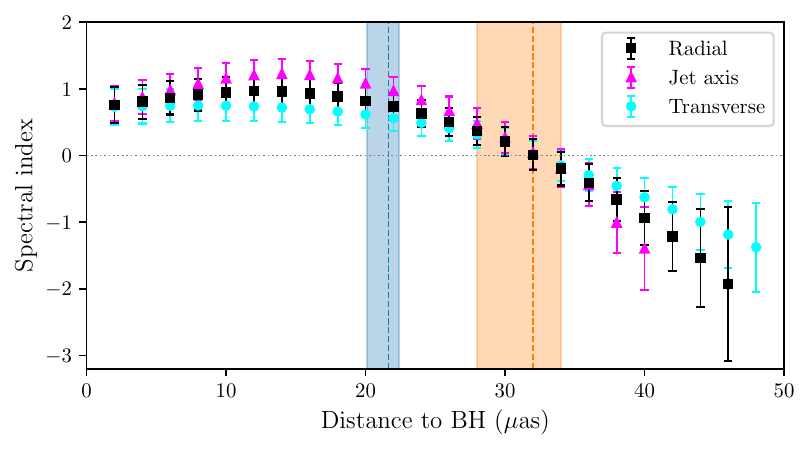}
    \caption{
    Radial spectral index profiles derived from the 86 and 230 GHz images. The black curve shows the azimuthally averaged profile, while the magenta and cyan curves represent profiles extracted along the jet axis (PA=$288^\circ$) and the transverse direction (PA=$18^\circ$), respectively. Vertical orange and blue lines indicate the characteristic ring radii at 86\,GHz ($32^{+2}_{-4}\,\mu$as) and 230\,GHz ($21.7^{+0.8}_{-1.6}\,\mu$as), with shaded bands showing uncertainties. The horizontal dotted line marks $\alpha = 0$.}
    \label{fig:spectral_index_radii}
\end{figure*}

In Figure~\ref{fig:spectral_index_map}, we show the spectral-index map (left) and the corresponding uncertainty map of M87* (right). 
The spectral-index distribution exhibits an overall ring-like morphology, broadly consistent with the structure seen in the total intensity images. The map shows more inverted (positive) spectral indices interior to the bright ring, while regions outside the ring are characterized by progressively steeper spectra. The transition between these regimes is spatially coincident with the ring scale.

We note a localized region of comparatively elevated spectral index along the north-western segment of the ring. This feature spatially coincides with the inferred foot point of the northern rim of the edge-brightened jet. If intrinsic, it may indicate locally enhanced synchrotron self-absorption or changes in the physical conditions of the emitting plasma near the jet–disk interface or the base of the emerging jet. However, given the limited $(u,v)$ coverage and the non-simultaneity of the 86 and 230\,GHz observations, the significance of such localized azimuthal variations remains uncertain. Future simultaneous multi-frequency VLBI observations will help assess the reality and physical origin of this feature.

The corresponding uncertainty map shows that the spectral-index uncertainty is relatively uniform across most of the central emission region, with typical values of $\sim$0.2. The uncertainty increases toward the edge of the image, reaching $\sim$0.5 in the outermost regions, where the spectral-index map exhibits its lowest values (down to $\sim -1.5$).

Figure~\ref{fig:spectral_index_radii} shows the radial spectral-index distribution derived from annular flux densities. Within the 1.3\,mm ring radius ($\sim$22\,$\mu$as), the spectral index remains nearly constant with positive values close to unity, indicating relatively inverted or optically thick synchrotron emission. A marginal enhancement is visible at $\sim 15\,\mu$as. Beyond the 1.3\,mm ring radius, the spectral index decreases steadily with increasing distance from the black hole. Near the 3.5\,mm ring radius ($\sim$32\,$\mu$as), it approaches zero, corresponding to a nearly flat spectrum, and continues to steepen at larger radii. 

This systematic radial evolution indicates a transition in the emission regime across the ring structure, consistent with optically thick plasma dominating the inner region and progressively optically thinner synchrotron emission at larger radii. The transition occurs on horizon scales and directly traces changes in plasma conditions in the immediate vicinity of the black hole. The coincidence of this spectral transition with the 86 GHz ring scale suggests that the observed 86 GHz ring is associated not only with a brightness enhancement but also with a change in the radiative properties of the emitting plasma.

For comparison, Figure~\ref{fig:spectral_index_radii} also shows spectral-index distributions measured along the jet-axis (PA=288$^\circ$) and transverse (PA=18$^\circ$) directions. 
Both directional distributions reproduce the same overall radial behavior as the radial profile, including the marginal enhancement near $\sim$15\,$\mu$as and the subsequent decline beyond the 1.3\,mm ring radius. \h{Within the 86\,GHz ring radius, the spectral-index profile along the jet direction is slightly higher than that along the transverse direction, in agreement with the spectral-index map (Fig.~\ref{fig:spectral_index_map}).} The overall agreement among the radial and directional distributions indicates that the radial trend is intrinsic rather than an artifact of image reconstruction.

\section{Discussion and summary}
We have presented the first spatially resolved spectral index measurements between 3.5\,mm and 1.3\,mm on horizon scales. The resulting spectral-index map exhibits a ring-like morphology, and a pronounced radial gradient is detected within $\sim50\,\mu$as of the black hole: the spectral index increases with radius in the inner region of the ring-like structure, reaches a nearly flat maximum slightly inside $\sim20\,\mu$as (coincident with the 1.3\,mm ring radius), and then decreases systematically at larger radii. The spectral index transitions from positive values in the inner region to negative values farther out, with the sign change occurring at $\sim 30\,\mu$as, close to the 3.5\,mm ring radius. This radial evolution provides a direct observational probe of the plasma conditions and radiative regime near the event horizon, tracing the transition from optically thick emission in the inner region to optically thinner synchrotron emission at larger radii.

For comparison, we also consider the spatially unresolved spectral index. Based on the 2018 images at 3.5\,mm and 1.3\,mm, the integrated spectral index within the compact region ($<100\,\mu$as) is $\alpha \simeq -0.04$, consistent with a nearly flat spectrum. 
This is in agreement with earlier spatially unresolved, multi-wavelength measurements obtained during the EHT 2018 campaign \citep{2024A&A...692A.140A}. Quasi-simultaneous VLBI observations with the East Asian VLBI Network (EAVN) and Korean VLBI Network (KVN) between 1.3\,cm (22\,GHz) and 2.3\,mm (129\,GHz) during the EHT 2018 campaign constrained a spatially unresolved spectral index of $\alpha$ to lie within $-$($0.11$--$0.17$) for a larger compact region \citep[$<500\,\mu$as; see Appendix F2 in][]{2024A&A...681A..79E}. In addition, a previous study of the centimeter- to millimeter-wavelength synchrotron spectrum of the M\,87 core on $\lesssim0.8$\,mas scales, based on four years of fully simultaneous multi-frequency VLBI observations with KVN, found a core spectral index of $\alpha \gtrsim -0.37$ between 1.3\,cm (22\,GHz) and 2.3\,mm (129\,GHz) \citep{2018A&A...610L...5K}. These measurements are broadly consistent with the nearly flat integrated spectral index derived from the 3.5\,mm and 1.3\,mm images, indicating that the horizon-scale core emission dominates the VLBI spectrum, while the larger-scale regions contain contributions from optically thin jet emission.

We note that ALMA fluxes across the EHT sub-bands exhibit a steeper spectral index with $\alpha \approx –1.2$~\citep{2021ApJ...910L..14G}. This may reflect either optically thin emission from the extended jet on arcsecond scales or the onset of optically thin conditions in the compact horizon-scale core, though the current ALMA beam cannot spatially separate these contributions. If the steep spectrum arises from the compact core itself becoming optically thin, then the horizon-scale spectrum must exhibit curvature or a turnover between 3.5\,mm and 1.3\,mm. To directly probe this possibility, we have proposed the first VLBI observations of M87 with the GMVA at 110\,GHz, which will bridge the current frequency gap and provide a critical test of the core spectral shape on horizon scales.

The main radially dependent structure obtained in this work, i.e., a gentle rise in the spectral index toward the center followed by a systematic decline at larger radii, can be understood as a direct consequence of the more compact and sharply defined emission ring at 1.3\,mm compared to the broader emission ring at 3.5\,mm. This difference reflects frequency-dependent synchrotron opacity in the inner accretion flow: higher-frequency emission becomes optically thin closer to the black hole and thus probes more compact regions, whereas lower-frequency emission remains optically thick over a larger spatial extent. As a result, the inner region is increasingly dominated by higher-frequency emission, producing positive spectral indices, while the growing contribution of lower-frequency emission at larger radii leads to a gradual steepening of the spectrum. The transition from positive to negative spectral index near the 3.5\,mm ring radius is consistent with a change from partially optically thick to predominantly optically thin synchrotron emission. 

From a theoretical perspective, independent of specific simulation assumptions, the supermassive black hole is expected to be surrounded by a radiatively inefficient accretion flow (RIAF), which is a hot, geometrically thick, and optically thin disk with a low Eddington ratio \citep{2014ARA&A..52..529Y}. Under ideal (infinite) resolution, theory predicts a pronounced spectral-index enhancement near the photon ring, where the emission originates from the most extreme conditions of gravitational lensing, magnetic fields, and plasma environment \citep{2023MNRAS.519.4203R,2025JKAS...58...17C,2019MNRAS.486.2873C}. At 84-86\,GHz, the photon-ring region is predicted to exhibit $\alpha>2$ together with a flip in the linear polarization direction due to optically thick nonthermal emission \citep{2024PASJ...76.1211T}, whereas at 214.1--228.1 GHz, the emission is optically thin and the predicted spectral index is $\alpha\sim-0.5-0$ \citep{2023MNRAS.519.4203R}.

A qualitatively similar radial decline to that observed has been reported in recent theoretical work by \citet{2023MNRAS.519.4203R}, who applied general relativistic radiative transfer to GRMHD simulations of M87 for a wide range of black hole spins, accretion states, electron thermodynamic prescriptions and electron distribution functions. They found that the radial spectral-index gradient is a generic feature and that its slope can provide a useful diagnostic of the underlying plasma properties. Our measured 86-230\,GHz profile exhibits a relatively steep gradient of $d\alpha/dr\sim-0.11\ \mu{\rm as}^{-1}$ in the declining region ($\gtrsim20\ \mu$as), comparable to the steepest thermal-electron Standard and Normal Evolution (SANE) models in their study (profiles at 214.1--228.1 GHz). Some nonthermal Magnetically Arrested Disks (MAD) models with high spin and small $\rm{R_{high}}$ also exhibit relatively steep gradients and unresolved spectral indices closer to our measured value ($\alpha \simeq -0.04$). However, this comparison should be regarded as indicative only, given the substantially different frequency coverage and modeling assumptions.


In these simulations, the photon-ring spectral-index spike is largely smoothed out once realistic observational resolution is applied, while the overall radial trend is preserved. This behavior is consistent with our measurements, which reveal a smooth radial decline and only a marginal enhancement in the inner region, suggesting that the radial spectral index gradient is a robust signature of horizon-scale emission, whereas fine-scale features associated with the photon ring remain challenging to detect with current observational capabilities.

Previous spatially resolved spectral index measurements at lower frequencies primarily trace the jet rather than the jet base or accretion region. \citet{2023A&A...673A.159R} reported that the 1.3/0.7\,mm (22/43\,GHz) spectral index decreases from $\sim -0.2$ near the jet core \citep[with the 43\,GHz core located at $\sim41\,\mu$as from the black hole center;][]{2011Natur.477..185H} to $\sim -2.5$ at a distance of $\sim 6$\,mas. A similar trend is seen in the 24/43\,GHz spectral index map from the EHT 2018 campaign \citep[Fig.~15 in][]{2024A&A...692A.140A}. When combined with existing 22/43\,GHz measurements and forthcoming 43/86\,GHz spectral mapping, as well as future 230/345\,GHz observations, these datasets will enable increasingly continuous constraints on M\,87*’s spectral properties from event-horizon scales to parsec-scale jet structures.

Spatially resolved spectral-index measurements also provide new constraints on the physics of the inner accretion flow. They can help distinguish between accretion scenarios such as MAD and SANE, constrain the electron energy distribution, and inform estimates of physical parameters such as black hole spin, inclination, and magnetic field strength. In particular, characteristics of the radial spectral-index profile, such as the location of spectral transitions and the radial gradient, together with the azimuthal distribution in spectral-index map may serve as effective observables that help reduce degeneracies in theoretical models and provide a useful diagnostic of accretion and emission processes.

Following the 2018 observations, the EHT and GMVA have continued high-resolution imaging of M87* at 1.3 and 3.5\,mm. The next-generation EHT (ngEHT) will further extend these efforts to 345\,GHz, with additional stations, wider bandwidths, and higher recording rates \citep{Johnson2023Galax11.61,Raymond2024AJ168.130}. A key challenge for spectral studies remains the intrinsic resolution mismatch between frequencies, which could be mitigated by space-VLBI extending baselines beyond Earth. RadioAstron has already imaged M87 at 1.3\,cm (22\,GHz) with $\sim150~\mu$as resolution \citep{Kim2023ApJ952.34}, and future missions such as the Black Hole Explorer (BHEX), the Space-based High-resolution Array for Radio astronomy and Physics (SHARP)/Event Horizon Imager (EHI), and Millimetron aim to resolve the photon ring at sub-millimeter wavelengths \citep{Johnson2024SPIE13092E..2DJ,Roelofs2019AA625.124,Likhachev2022MNRAS511.668}. In parallel, advances in multi-frequency imaging algorithms \citep[e.g.,][]{2023ApJ...945...40C} will further improve the accuracy and robustness of spectral index maps.

In summary, our spatially resolved spectral index measurements reveal a robust radial gradient on horizon scales, characterized by a gentle rise toward the black hole followed by a systematic decline at larger radii. This behavior reflects frequency-dependent synchrotron opacity in the inner accretion flow and is consistent with GRMHD-based theoretical predictions, even when realistic observational resolution is taken into account. When combined with lower-frequency measurements tracing the jet, these results provide a nearly continuous view of M87*’s spectral properties from the event horizon to parsec scales. Collectively, our findings offer the first observational evidence for a systematic radial spectral transition on horizon scales in M87* and demonstrate that spectral-index mapping can serve as a powerful probe of accretion physics, jet launching, and plasma conditions in the immediate vicinity of a supermassive black hole, a capability that will be further enhanced by future multi-frequency, high-resolution VLBI observations.

\begin{acknowledgments}
We thank the anonymous reviewer for the constructive comments and helpful suggestions, which have significantly improved our paper. We thank T. P. Krichbaum and J.-Y. Kim for helpful comments. This work is supported by the National Science and Technology Major Project of China (2024ZD1100601), the National Natural Science Foundation of China (Grant No. 12325302, 11933007), the Key Research Program of Frontier Sciences, CAS (grant no. ZDBS-LY-SLH011), and the Shanghai Pilot Program for Basic Research, Chinese Academy of Sciences, Shanghai Branch (JCYJ-SHFY-2021-013). This paper makes use of the following ALMA data: ADS/JAO.ALMA\#2017.1.00842.V and ADS/JAO.ALMA\#2017.1.00841.V. ALMA is a partnership of ESO (representing its member states), NSF (USA) and NINS (Japan), together with NRC (Canada), MOST and ASIAA (Taiwan), and KASI (Republic of Korea), in cooperation with the Republic of Chile. The Joint ALMA Observatory is operated by ESO, AUI/NRAO and NAOJ. 
\end{acknowledgments}


\facilities{ALMA, EHT, GMVA}

\software{astropy\citep{2013A&A...558A..33A,2018AJ....156..123A,2022ApJ...935..167A}}

\appendix

\section{Discussion of Uncertainties}
In this Appendix, we discuss the primary sources of uncertainty affecting the spatially resolved spectral index measurements presented in this work. These include systematic effects related to image alignment (Appendix~\ref{sec:A_alignment}), resolution matching between frequencies (Appendix~\ref{sec:A_beam_size}), and absolute flux density scale uncertainties (Appendix~\ref{sec:A_flux_density}). Our aim is to quantify the impact of these effects on the inferred radial spectral index profiles and to demonstrate the robustness of the detected radial trends.

\subsection{Uncertainties from image alignment}
\label{sec:A_alignment}
Accurate image alignment is a critical requirement for reliable spatially resolved spectral index measurements. In this work, we align the 3.5\,mm (86G\,Hz) and 1.3\,mm (230\,GHz) images by assuming that the centers of the ring-like structures coincide and correspond to the black hole position. Here we evaluate the validity of this assumption and discuss its implications for the inferred spectral index profile.

At centimeter wavelengths, VLBI observations commonly reveal a frequency-dependent shift of the apparent core position of the jet (the so-called core shift). Based on measurements at frequencies $\leq 43$\,GHz, \citet{2011Natur.477..185H} reported a core-shift relation of $r \propto \nu^{-0.94}$ for M\,87, consistent with expectations for a conical jet. A direct extrapolation of this relation to millimeter wavelengths would suggest a positional offset of order $\sim12\,\mu$as between 86 and 230\,GHz.

However, such an extrapolation is unlikely to be valid in the horizon-scale regime. VLBI observations have shown that the M87 jet exhibits a parabolic geometry over a wide range of spatial scales, extending from sub-parsec scales down to the jet-launching region \citep[e.g.,][]{2012ApJ...745L..28A}. More recently, the 86 GHz observations of M87* presented by \citet{2023Natur.616..686L} directly resolve this parabolic jet structure on horizon scales, deviating substantially from the conical jet assumption adopted at lower frequencies. This morphological difference implies that the core-shift relation derived at centimeter wavelengths cannot be straightforwardly extended to 86/230 GHz. 

More importantly, at both 3.5\,mm and 1.3\,mm the dominant emission forms a resolved ring-like structure centered on the black hole. This structure is widely interpreted as arising from the inner accretion flow and strong gravitational lensing, rather than from a compact, self-absorbed jet core. In this regime, the jet-based core-shift framework, defined as a frequency-dependent displacement of the VLBI core, is therefore not physically applicable.

Furthermore, the resolved images at both 86 and 230 GHz exhibit asymmetric brightness distributions around the black hole. Such asymmetries likely reflect Doppler boosting, magnetic-field structure, or intrinsic variability in the accretion flow. Consequently, the location of any unresolved jet contribution cannot be used as a reliable reference for image registration.

Given these considerations, aligning the 86 and 230\,GHz images using the center of the resolved ring structure provides the most robust and physically motivated approach for this analysis. This choice minimizes systematic alignment errors and ensures that the derived spectral index profiles are referenced to a common gravitational center.

\subsection{Uncertainties from convolving beam size}
\label{sec:A_beam_size}
When computing the spatially resolved spectral index, it is essential to use images with matched angular resolution across frequencies. For the images analyzed in this work, the typical resolution is $\sim 20~\mu$as at 1.3\,mm and $\sim 79 \times 37~\mu$as (full width at half maximum, FWHM, position angle $-63^\circ$) at 3.5\,mm, as indicated by the corresponding CLEAN beams. Convolving the 1.3\,mm image with a beam as large as the 3.5\,mm beam (even along the minor axis) would eliminate the ring-like structure at 1.3\,mm, whereas convolving the 3.5\,mm image to 20\,$\mu$as resolution would impose unrealistically high super-resolution, beyond the intrinsic capabilities of the 3.5\,mm observations.

\begin{figure*}
    \centering
    \includegraphics[width=0.7\linewidth]{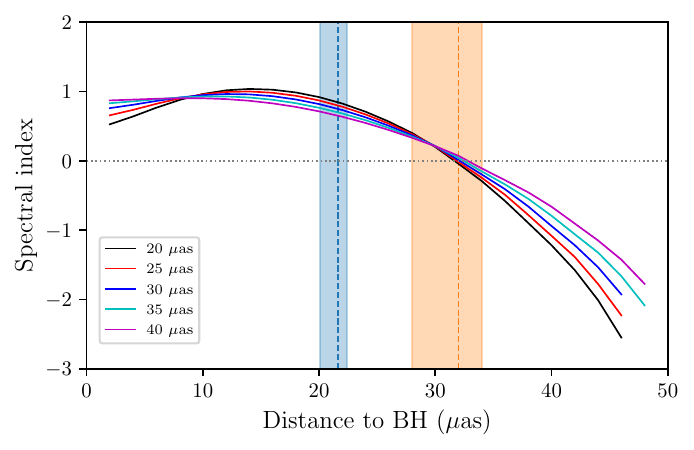}
    \caption{Spectral index radial profiles computed from images convolved with circular Gaussian beams of different sizes (FWHM = 20, 25, 30, 35, and 40\,µas), using the GMVA 2018 3.5\,mm and EHT 2018 1.3\,mm images. This figure illustrates the impact of beam size on the radial spectral index, highlighting how intermediate smoothing preserves the ring morphology while enabling a robust comparison between the two frequencies.}
    \label{fig:beam_size_radii}
\end{figure*}

\begin{figure*}
    \centering
    \includegraphics[width=\linewidth]{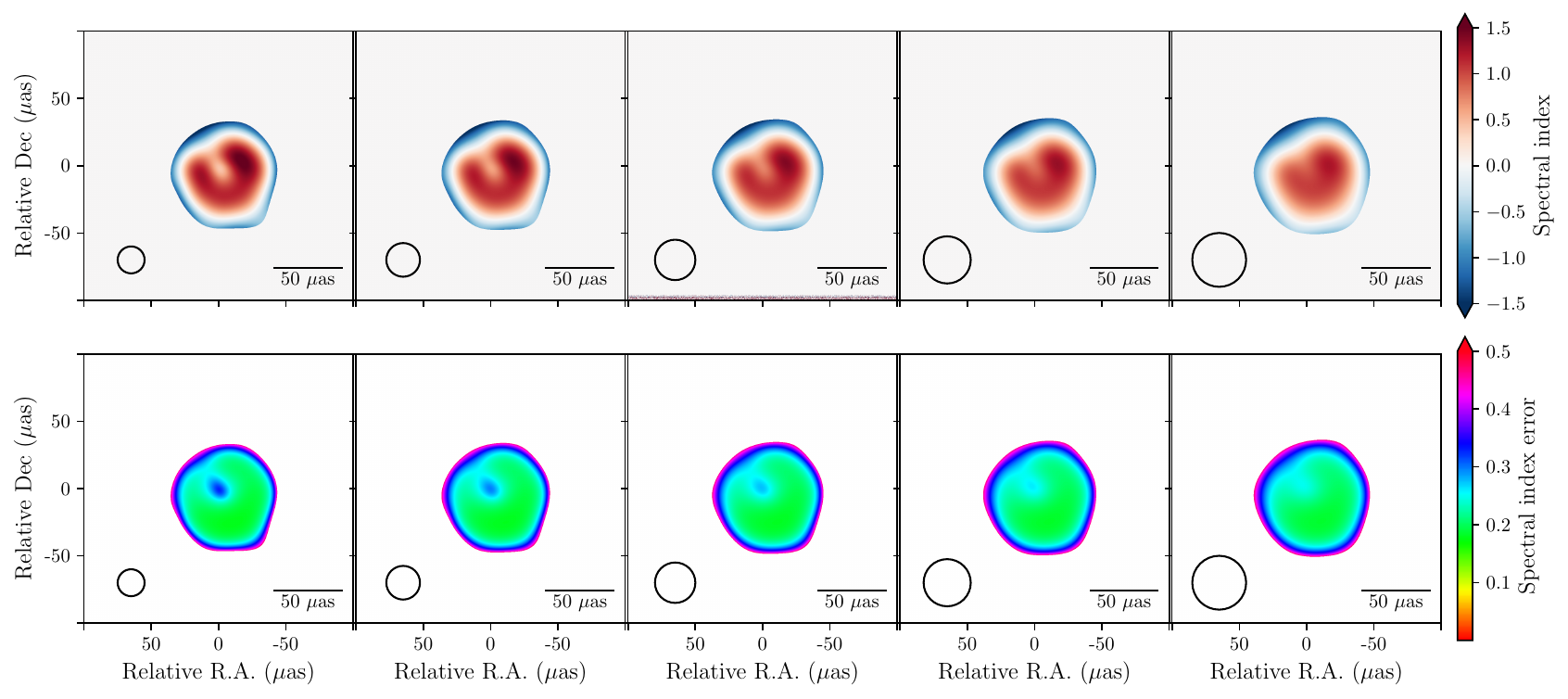}
    \caption{Spectral index maps (top row) and their corresponding error distributions (bottom row) derived from the GMVA 2018 3.5\,mm and EHT 2018 1.3\,mm images. Columns show results for different circular Gaussian beam sizes (FWHM = 20, 25, 30, 35, and 40\,µas), with the beam illustrated as a circle in the lower-left corner of each panel. All maps are masked to include only pixels where the intensity in both bands exceeds $3\,\sigma_{\rm rms\nu}$.}
    \label{fig:beam_size_map}
\end{figure*}

To balance these considerations, we convolved both images to a common circular Gaussian beam, testing FWHM values of 20, 25, 30, 35, and 40\,$\mu$as. Figures~\ref{fig:beam_size_radii} and \ref{fig:beam_size_map} show the resulting spectral index radial profiles and two-dimensional maps. At 40\,$\mu$as, the radial profile becomes monotonically decreasing and the map nearly featureless, while at 20\,$\mu$as, the higher-frequency 1.3\,mm structure dominates over the 3.5\,mm image.

A beam size of 30\,$\mu$as provides the best compromise as it preserves the ring morphology, maintains meaningful radial gradients, and ensures a balanced representation of the emission structure at both frequencies. This beam was adopted for all spectral index maps and profiles presented in this work.

We note that these spectral index profiles are derived from images convolved with a finite beam, which can influence the apparent slope. Observed gradients should therefore be interpreted with the beam resolution in mind, and meaningful comparisons with theoretical or simulated models should be performed after convolving the models to the same resolution.


\subsection{Uncertainties from compact flux density}
\label{sec:A_flux_density}

The accuracy of the spectral index measurements depends on both the amplitude calibration of the VLBI arrays and the compact flux density of the horizon-scale core. For GMVA 86\,GHz observations, the amplitude calibration uncertainty is typically estimated to lie between roughly 15\,\% and 30\,\% \citep{2018A&A...610L...5K,2023Natur.616..686L}, while for the EHT 230\,GHz observations, the amplitude calibration uncertainty is approximately 15\,\%~\citep{2024A&A...681A..79E}. Propagating these uncertainties to the spectral index measurement implies a maximum systematic uncertainty of $\Delta\alpha \lesssim 0.3$. Because amplitude calibration uncertainties act primarily as multiplicative scaling factors applied uniformly to the images at each frequency, they mainly shift the absolute value of the spectral index while leaving the spatial structure and the radial trends discussed in this work largely unchanged.

For the spatially resolved spectral index of the compact horizon-scale core ($<100,\mu$as), the dominant source of systematic uncertainty is the compact flux density itself. For the 230\,GHz EHT image, the lack of short baselines limits the direct determination of the compact flux from the visibilities, giving a range of 0.3--1.13\,Jy. Using quasi-simultaneous multi-wavelength VLBI measurements and extrapolations, this range is narrowed to 0.5--0.7\,Jy \citep[][]{2024A&A...681A..79E}, and we adopt $F_{\rm cpct,230} = 0.58$\,Jy for the fiducial image. For the 86\,GHz GMVA image, the presence of short baselines allows a tighter constraint of the compact flux, yielding 0.5--0.6\,Jy from CLEAN and SMILI reconstructions~\citep{2023Natur.616..686L}; we adopt $F_{\rm cpct,86} = 0.56$\,Jy. Geometric model fitting of the jet-subtracted visibility data gives a zero-baseline flux of 0.58\,Jy, consistent with this choice \citep{2023Natur.616..686L}.

\begin{figure*}
    \centering
    \includegraphics[width=\linewidth]{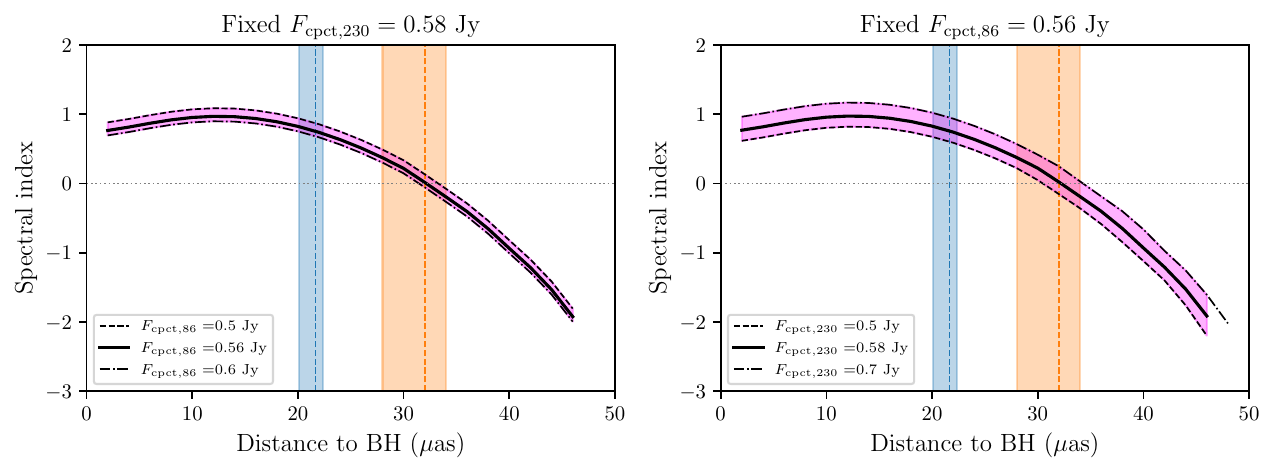}
    \caption{Spectral index radial profiles derived from the GMVA 2018 and EHT 2018 images for different assumptions of compact flux density. All images have been convolved with a 30 $\mu$as beam. \textbf{Left panel:} Profiles with a fixed compact flux density of $F_{\rm cpct,230}=0.58$ Jy and varying $F_{\rm cpct,86}$ from 0.5 Jy (black dashed line) to 0.6 Jy (black dash–dotted line); the spanned range is indicated by the magenta shaded region. The case with $F_{\rm cpct,86}=0.56$ Jy and $F_{\rm cpct,230}=0.58$ Jy, adopted in this work, is shown by the thick black solid line. \textbf{Right panel}: Same with the left panel, but for profiles with a fixed $F_{\rm cpct,86}=0.56$ Jy and varying $F_{\rm cpct,230}$ from 0.5 Jy (black dashed line) to 0.7 Jy (black dash–dotted line). }
    \label{fig:uncertainties_flux_density}
\end{figure*}

Fig.~\ref{fig:uncertainties_flux_density} illustrates how variations in $F_{\rm cpct,86}$ and $F_{\rm cpct,230}$ affect the radial spectral index profiles. While the absolute values of $\alpha$ shift slightly with different flux assumptions, the overall radial shape, the location of the transition from positive to negative spectral index, and the modest peak near the 230\,GHz ring remain robust. This demonstrates that the main radial trends of the spectral index are primarily determined by the compact flux of the horizon-scale core, rather than by amplitude calibration uncertainties.

\bibliography{M87}{}
\bibliographystyle{aasjournalv7}

\end{document}